\def\aa{{A\&A}}

\def\aj{{AJ}}
\def\annrev{{ARA\&A}}
\def\apj{{ApJ}}
\def\apjs{{ApJS}}

\def\mnras{{MNRAS}}

\documentclass[cup5b]{caps}
\usepackage{graphicx}
\usepackage{amssymb}
\usepackage{ociwsymp3e}  

\HeadText{Neal A.\ Miller}  

\newcommand\OII{[O\,{\sc ii}]}

\begin{document}

\pagenumbering{arabic}

\author[]{Neal A.\ Miller\\NRC \& NASA GSFC, Greenbelt, MD}

%
%

\chapter{Star Formation in \\ Cluster-Cluster Mergers}

\begin{abstract}

Radio continuum emission at 1.4 GHz was used to identify active galaxies in a sample of 20 nearby Abell clusters. The radio emission indicates either an active galactic nucleus or current star formation, and optical spectroscopy was used to evaluate which of these dominates for each of the radio-selected galaxies. In an analysis which parallels the blue fractions of Butcher-Oemler studies, we calculated radio galaxy fractions for the clusters. One cluster in particular shows a dramatic increase in activity relative to the others: the cluster-cluster merger Abell 2255. We compare the results for Abell 2255 with those for Abell 2256 to assess the role of cluster-cluster mergers on the star formation activity of member galaxies. From these clusters, as well as several identified in other studies, a picture emerges in which substructure and cluster dynamical activity are of great importance in understanding evolutionary phenomenon such as the Butcher-Oemler Effect. Increased fractions of active galaxies (be they the blue galaxies of Butcher-Oemler studies or radio galaxies as discussed here) naturally result as clusters are built through mergers of smaller groups in hierarchical formation scenarios. The observed spread in fractions of active galaxies at any given redshift reflects the spread in cluster dynamical state.

\end{abstract}

\section{Introduction}

The exciting evidence for evolution in distant clusters of galaxies \cite{bo1,bo2,morphs1,morphs2} has diverted attention from nearby clusters. This is unfortunate, since although nearby clusters show less evidence of galaxy activity they may be studied to much greater depth. The findings may then be applied to studies of higher redshift clusters and to the general understanding of the cause for evolution in cluster populations with redshift.

The potential importance of dust in the active galaxies of distant clusters \cite{morphs2,smail} underscores the need for selecting samples independent of the effects of dust extinction. Radio continuum emission (in this study we will use a frequency of 1.4 GHz) is a powerful probe of activity caused by both active galactic nuclei (AGN) and current star formation \cite{condon}. It is particularly attractive for this latter type of activity, as the radio continuum emission is directly related to the star formation rate (SFR). Consequently, a radio flux-limited sample drawn from a given cluster will select all AGN and galaxies forming stars at a rate greater than some adopted threshold.

We have constructed a sample of radio galaxies in 20 nearby Abell clusters \cite{sample,spec}. For the nearer clusters, the radio data are taken from the NRAO VLA Sky Survey \cite{nvss} while for the more distant clusters we have collected our own VLA data \cite{myA2255,myA2256}. The luminosities associated with the flux limits of these data correspond roughly to that of the Milky Way, meaning our completeness limit is a SFR of 4M$_\odot$ yr$^{-1}$ (for stars with masses $0.1 - 100$ M$_\odot$, using the relationship of \cite{yrc}). The more sensitive observations (i.e., the nearest NVSS clusters and our VLA data for the more distant clusters) detect SFRs as low as 1M$_\odot$ yr$^{-1}$. Velocity measurements based on optical spectra indicate whether the radio galaxies are members of their respective clusters, or merely seen in projection. The full sample contains over 400 cluster radio galaxies and over 150 foreground and background radio galaxies. We have quality optical spectra for about half of the cluster radio galaxies, which enables us to identify the source of the radio emission as either star formation or an AGN.

Using this sample, we ask the question: Do any clusters show more evidence for activity and if so, why? The motivation is to understand the Butcher-Oemler effect, including both the trend and its scatter. As noted, we also wish to establish a nearby benchmark for evolutionary studies of distant clusters.

\section{Results}

Analogous to the blue fraction of Butcher-Oemler studies, we calculated the fraction of cluster galaxies associated with radio emission ($f_{RG}$). For this calculation, we used all galaxies brighter than $M_R = -20$ (determined from the systemic velocities of the clusters). The optical spectroscopy indicated which radio galaxies were cluster members and which were foreground/background objects, while the total number of galaxies for each cluster was corrected for estimated background counts. To place the clusters on even footing, we used only sources with radio luminosities greater than the NVSS completeness limit for the most distant cluster which used NVSS data (i.e., radio detections at lower luminosities were considered non-detections for the fractional testing). In each cluster, we surveyed from the cluster center out to a radial distance of 3$h_{75}^{-1}$ Mpc, or $\sim1.5R_{Abell}$. For the $f_{RG}$ calculations we applied a radial limit of 2$h_{75}^{-1}$ Mpc in order to reduce variation caused by the background correction.

\begin{figure}
    \centering
    \includegraphics[width=11cm,angle=0]{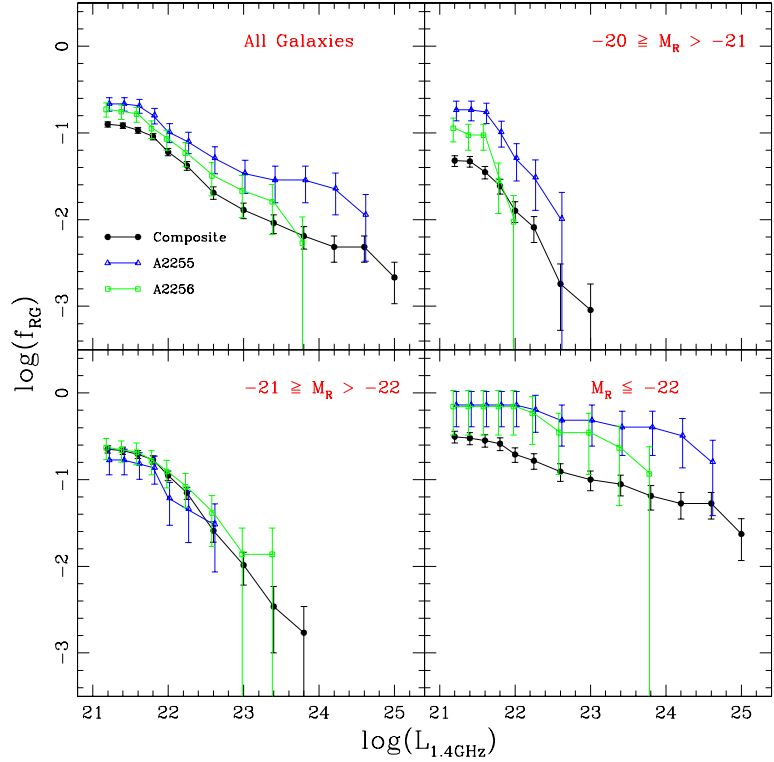}
    \caption{Cumulative RLF, plotted using three different cuts based on optical magnitude (the total of these is shown at top left). The composite sample, in black, includes 18 nearby Abell clusters. Abell 2255 is depicted in blue, while Abell 2256 is shown in green. The radio galaxy fractions discussed in the text are based on galaxies with radio luminosities greater than $\log (L_{1.4}) = 21.84$, so the first three bins in the plots do not contribute.}
    \label{fig-comprlf}
  \end{figure}

The $f_{RG}$ values were then compared via a $\chi^2$ test. Two clusters showed an enhancement in activity (larger $f_{RG}$ than the other clusters of the sample, significant at $>99\%$ confidence), Abell 1185 and Abell 2255. To explore what types of galaxies cause these enhancements, we investigated $f_{RG}$ as a function of optical magnitude. Three bins in optical magnitude were created: $M_R \leq -22$, $-21 \geq M_R > -22$, and $-20 \geq M_R > -21$. For reference, $M^*_R = -22$ for ellipticals \cite{owenlaing} so these approximately correspond to bright ellipticals, larger spirals and S0s, and later-type spirals and irregulars, respectively. The increased $f_{RG}$ for Abell 1185 was revealed to be caused by the intermediate optical magnitude galaxies. For Abell 2255, the increased $f_{RG}$ was caused by both the optically brighter and fainter galaxies (see Figure \ref{fig-comprlf}). In fact, the excess of the fainter galaxies was significant at over $99.9\%$. This result is robust against variations in the background count. In fact, should {\it all} optically-faint galaxies in the direction of Abell 2255 be cluster members (hence no background correction and a minimum value for $f_{RG}$), the radio galaxy enhancement would still be significant at $97.5\%$ confidence.

With complete optical spectroscopy for Abell 2255, we further analyzed the emission source for the radio galaxies. Of the 23 optically-faint galaxies in this cluster, 21 had radio emission associated with star formation. The other two were weak emission-line AGN. Among the star-forming galaxies there were many examples of rigorous star formation, including three starbursts with EW(\OII )$>$ 40$\mbox{\AA}$ and four dusty starbursts with slightly weaker EW(\OII ) but strong H$\delta$ absorption (see \cite{morphs1,morphs2}). Additionally, one galaxy had a strong post-starburst spectrum. Thus, there is strong evidence for current and dramatic activity within the galaxy population of Abell 2255.

\section{Discussion}

Why is Abell 2255 so special? We searched for correlations between $f_{RG}$ and parameters such as cluster richness and overall galaxy distribution (i.e., is the cluster compact or fairly dispersed?), but found none. In studies of more distant Butcher-Oemler clusters, an increased $f_{RG}$ for Abell 2125 as compared to Abell 2645 was noted despite the nearly identical redshift and richness of these clusters \cite{dwarka,owen}. The authors concluded that the discrepancy was likely the result of Abell 2125 being a cluster-cluster merger rich in substructure whereas Abell 2645 is more virialized. This explanation is attractive in the case of Abell 2255, as it is also a strong merger candidate (e.g., \cite{burns,feretti,hill}).

The main problem with this explanation is that other clusters in our sample are also probable mergers. As a case example, we discuss Abell 2256 (see \cite{berr,sun} for merger evidence). Radio observations of this cluster were performed concurrently with those of Abell 2255, and the two clusters are also comparable in richness and redshift. However, with the possible exception of the optically-brightest galaxies the $f_{RG}$ values for Abell 2256 were normal (see Figure \ref{fig-comprlf}). Consistent with this result is the measured blue fraction for Abell 2256, $f_B = 0.03 \pm 0.01$ \cite{bo2}.

The key appears to be a combination of the merger stage and large-scale structure. We assessed the substructure in each cluster using a variety of techniques, including the Dressler-Schectman test \cite{ds} and a KMM analysis to separate galaxies into respective substructures \cite{abz}. Each cluster showed evidence for substructure, but the specific distributions of the star-forming galaxies are illustrative. In Abell 2255, the star-forming galaxies were found in a strong North-South alignment which proved statistically significant through Monte Carlo testing. This alignment is perpendicular to the merger axis, which is East-West and fairly close to the plane of the sky \cite{burns}. We would expect such an arrangement if the cluster-cluster merger initiated bursts of star formation as galaxies experienced spikes in ram pressure as they crossed the shock front of a very recent merger \cite{roet}. Presumably, the precursors to these starbursts were field-like spirals at the peripheries of the pre-merger systems. In Abell 2256, the star-forming galaxies are clustered slightly to the North of the cluster center, and prove to be almost exclusively associated with a high velocity group composed of around 30 members. Interestingly enough, this substructure is not one of the culprits suggesting a large-scale merger in Abell 2256; many more galaxies are associated with the primary cluster and a large infalling subcluster, the two obvious merger partners from the X-ray \cite{sun}. That the star-forming galaxies originate in an infalling group underscores that the merger of a pair of rich clusters alone is likely not enough to cause an increase in the active galaxy population relative to other clusters. Some ``fuel'' is required, in particular a population of gas-rich galaxies which can undergo a starburst. These are apparently supplied by outlying groups and field galaxies which are thrown into the mix by the merger. Note that the main population responsible for the increased $f_{RG}$ in Abell 2255 are galaxies one to two magnitudes fainter than $M^*$, similar to the magnitudes of late-type field spirals and the blue galaxies observed in Butcher-Oemler studies \cite{couch}. In summary, we apparently see an enhancement in activity for Abell 2255 as a result of both timing (viewing the system after very recent core passage) and large-scale structure (having available galaxies which could undergo a starburst).

The relationship of this explanation to the Butcher-Oemler Effect is also quite attractive. Currently favored cosmological models indicate that clusters are actively being assembled from smaller clusters and groups since $z\approx 1$, with the rate of such mergers increasing with redshift. Consequently, the trend of increased blue fraction (and $f_{RG}$) with redshift is indicative of the assembly history of clusters. By $z\approx 0$, most clusters are fairly well assembled and virialized and consequently contain fewer active galaxies. The spread in the fraction of active galaxies at any given redshift is a reflection of the diversity of cluster dynamical states. More relaxed clusters would have fewer active galaxies, while clusters in the process of assembly would have more active galaxies. In fact, this explanation has been noted to alleviate the discrepancy in galaxy activity observed for the MORPHS and CNOC clusters, as the CNOC sample are X-ray selected and consequently more frequently associated with relaxed clusters \cite{cnoc}. Lastly, this spread in dynamical state indicates that galaxy activity may be observed even in nearby clusters.

Of course, this explanation requires a much larger sample of clusters to fill in necessary details. Similar results are being found for very rich clusters at higher redshift, in a sample which includes the aforementioned Abell 2125 \cite{glenn}. We have also obtained deep, wide-field radio and optical images of several nearby clusters (out to $z\sim0.05$) in a range of merger stages.

\vspace*{0.25in}

The author thanks the Carnegie Observatories for an informative and enjoyable conference, and Frazer Owen, Bill Oegerle, and John Hill for their involvement in this work. The author also acknowledges the National Radio Astronomy Observatory for a predoctoral appointment, and the National Research Council for postdoctoral support through an Associateship award held at NASA Goddard Space Flight Center.

\end{document}